\documentclass{aa}  
\usepackage{graphicx}
\usepackage{natbib}
\usepackage{txfonts}
\begin{document}
   \title{Spatial mapping of ices in the Oph-F core}

   \subtitle{A direct measurement of CO depletion and the formation of CO$_2$}

   \author{K. M. Pontoppidan \inst{1,2}
          }

   \offprints{K. M. Pontoppidan}

   \institute{California Institute of Technology, Division for Geological and Planetary Sciences, MS 150-21, Pasadena, CA 91125, USA\\
              \email{pontoppi@gps.caltech.edu}
\and
   Leiden Observatory, P.O. Box 9513, NL-2300 RA, Leiden, The Netherlands\\
              }

   \date{Received; accepted}

 
  \abstract
    {}
    {Ices in dense star-forming cores contain the bulk of volatile molecules apart from H$_2$ and thus represent a large fraction of dark
    cloud chemistry budget. Mm observations of gas provide indirect evidence for significant freeze-out of CO in the densest 
    cores. To directly constrain the freeze-out profile of CO, the formation route of CO$_2$ and the carrier of the
    6.8\,$\mu$m band, 
    the spatial distribution of the CO/CO$_2$ ice system and the 6.8\,$\mu$m band carrier are measured in a nearby dense core.}
   {VLT-ISAAC, ISOCAM-CVF and Spitzer-IRS archival mid-infrared (3-20\,$\mu$m) spectroscopy of young stellar objects is used to 
    construct a map of the abundances of CO and CO$_2$ ices in the Oph-F star-forming core, probing core radii from $2\times 10^3$ to 
    $14\times 10^3\,$AU or densities from $5\times 10^4$ to $5\times 10^5\,\rm cm^{-3}$
    with a resolution of $\sim 3000\,$AU.}
   {The line-of-sight averaged abundances relative to water ice of both CO and CO$_2$ ices increase monotonously with decreasing 
    distance to the core center. The map traces the shape of the CO abundance profile between freeze-out ratios of 5--60\%
    and shows that the CO$_2$ ice abundance increases by a factor of 2 as the CO freezes out. It is suggested that this indicates a formation
    route of CO$_2$ on a CO ice surface to produce a CO$_2$ component dilute in CO ice, in addition to a fraction of the CO$_2$ formed 
    at lower densities along with the water ice mantle. 
    It is predicted that the CO$_2$ bending mode band profile should reflect a high CO:CO$_2$ number ratio
    in the densest parts of dark clouds. In contrast to CO and CO$_2$, the abundance of the carrier of the 
    6.8\,$\mu$m band remains relatively constant throughout the core. A simple freeze-out model of the CO abundance profile is used to estimate
    the binding energy of CO on a CO ice surface to $\rm 814\pm 30 \,\rm K$. 
     }
   {}


   \maketitle
%

\section{Introduction}

Theory has long predicted that molecules freeze out onto dust grains in dense molecular clouds causing gas-phase abundances
to drop by orders of magnitude. A series of recent measurements of the gas-phase abundances of volatile molecules such as CO and N$_2$H$^+$
in dense cores have corroborated this conjecture \citep[e.g.][]{Caselli99, Bacmann02,Tafalla04}. CO ice has also been 
observed directly along
a growing sample of isolated lines of sight toward both embedded young stellar objects as well as background stars, 
showing that the CO molecules taken from the gas-phase re-appears in the solid phase \citep{Chiar94,Pontoppidan03a}. 
CO ice is an excellent tracer of 
freeze-out processes because it is the only abundant ice species known to form initially in the gas-phase before adsorbing to
a grain surface. 

\cite{Pontoppidan04} introduced the technique of ice mapping
at high ($\sim1000\,$AU) spatial resolution -- comparable to that of gas-phase maps. The procedure was demonstrated by 
constructing a map of the distribution of water and methanol ices in the outer envelope of the class 0 protostar SMM 4 in the
Serpens star-forming cloud. The ice map of SMM 4 was used to show that the water ice abundace remained constant over a relatively wide
range of densities, while the methanol abundance increased sharply by at least a factor of ten within 10\,000 AU of the center of SMM 4. 
In principle, ices are best mapped toward field stars located behind the cloud. However, such background stars are typically 
extremely faint in the mid-infrared wavelength region. Therefore, disk sources embedded in their parent molecular cloud core 
offer a convenient infrared continuum against which the ices in the core can be mapped, with appropriate caveats on the 
interpretation of the derived ice abundances. 
  
Direct mapping of CO ice abundances is highly
complementary to mapping of gas phase abundances, because ice maps are sensitive to depletion fractions as low as 5\%, while 
gas-phase abundance maps trace high depletion factors ($n({\rm ice})/n({\rm gas})$) of $\gtrsim 50\%$ \citep{Caselli99}. 
Furthermore, the depletion fraction as a function of gas density
in a core is strongly dependent on the properties of the dust grain surfaces. Direct maps of CO ice in a dense core 
therefore enables an independent measure of surface binding energies.

This letter presents the first spatial map of CO and CO$_2$ ices in a dense molecular core. {\it It will be demonstrated 
that by directly measuring the relation between ice abundances and gas densities, the following quantities can be constrained: 1)
the dependence of CO depletion on gas density with a large dynamical range, 2) the CO-CO binding energy, 3) the formation route
of CO$_2$ and the 6.8\,$\mu$m band carrier.}
The case study is the F core in the Ophiuchus molecular cloud complex (\cite{Motte98}). Embedded within this region, are at least
8--10 infrared-bright young stars. The lines of sight toward 5 of these young stars are used to probe the ices 
in front of each star in order to obtain a cross section of ice abundances across the minor axis of the core.


\section{Observations}

The individual spectra are taken from \cite{Dishoeck03} and \cite{Pontoppidan03a} of the 3.08\,$\mu$m and 4.67\,$\mu$m stretching modes 
of solid H$_2$O and CO, 
obtained with the Infrared Spectrometer And Array Camera (ISAAC) on the Very Large Telescope
(VLT) \footnote{Partly based on observations obtained at the European Southern Observatory, Paranal, Chile, within the observing programs 
164.I-0605 and 69.C-0441.}. The solid CO$_2$ component is probed along the same lines of sight using the InfraRed Spectrograph (IRS) on the 
Spitzer Space Telescope (AOR IDs 0009346048 and 0009829888 \footnote{This work is based in part on archival data obtained with the Spitzer 
Space Telescope, which is operated by the Jet Propulsion Laboratory, California Institute of Technology under a contract with NASA.}) 
\citep{Pontoppidan05, Lahuis06} 
as well as 5-16.3\,$\mu$m spectra obtained with ISOCAM-CVF (TDTs 29601715 and 29601813) 
\citep{Alexander03} \footnote{Partly based on observations with ISO, an ESA project with instruments funded by ESA Member States 
(especially the PI countries: France, Germany, the  Netherlands and the United Kingdom.)}. The Spitzer spectra were taken as part of the
Cores to Disks Legacy program \citep{Evans03}. 
Combining these facilities, high quality 3-16\,$\mu$m spectra are available for 5 sources within a radius of
15\,000\,AU from the center of the pre-stellar core, defined to be the 850\,$\mu$m peak of Oph-F MM2 ($\alpha$=16$^h$27 24.3, 
$\delta$=-24\degr 40 35, J2000) \citep{Motte98}. Only IRS 46 has no suitable spectrum between 5 and 10\,$\mu$m. 

\begin{figure*}
\includegraphics[width=18cm]{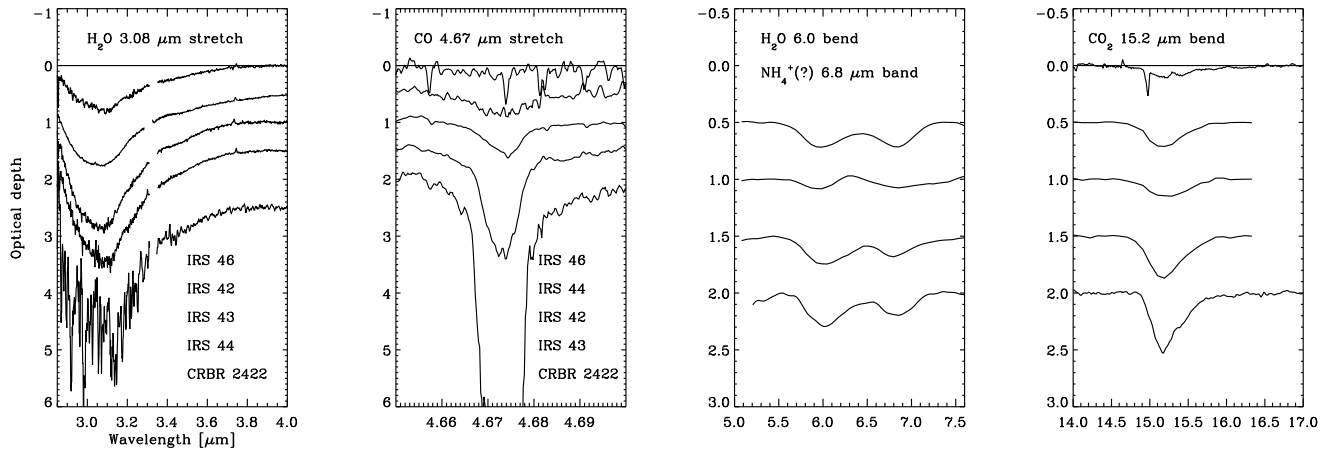}
  \caption{Overview of the H$_2$O, CO, 6.8\,$\mu$m and CO$_2$ ice bands (ordered left to right) used to construct the ice map of the 
Oph-F core. The 3.08\,$\mu$m
water ice bands shown on the left-most panel are ordered according to optical depth, for clarity, as indicated in the panel. In the remaining
panels, the sources are ordered according to projected distance to the core center, with the source furthest from the center (IRS 46) at the 
top. No 5-8\,$\mu$m spectrum is available for IRS 46. Note also that the 6.8\,$\mu$m band of IRS 42 is filled in by emission likely due to PAHs. 
All the spectra have been shifted vertically, for clarity.}
  \label{features}
\end{figure*}

\begin{figure}
\centering
\includegraphics[width=10cm]{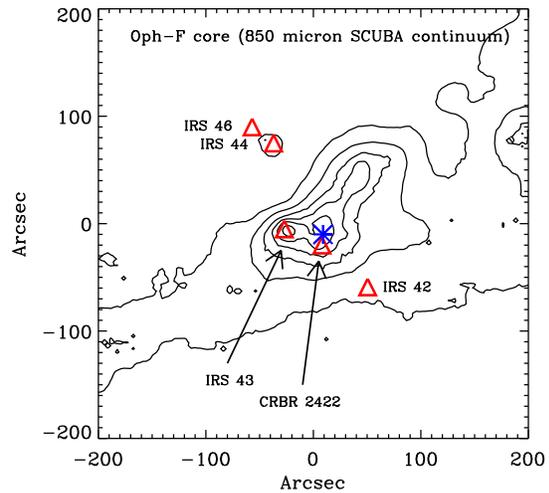}
  \caption{The locations (triangles) of the sources used to construct the ice map plotted on the SCUBA 850\,$\mu$m map. 
The star symbol indicates the center of the core. }
  \label{core_map}
\end{figure}

To determine the optical depths of the various ice bands continua were fitted using low-order polynomia. The CO and CO$_2$ bands 
are sufficiently narrow to make the continuum determination relatively straight-forward and unbiased. For the 3.08\,$\mu$m water ice band
a continuum is fitted to points between 3.8 to 4.0\,$\mu$m and a $K$-band photometric point from the 2MASS
catalogue. Finally, a local continuum between 5.5 and 8.0\,$\mu$m is used to extract optical depth spectra of the 6.8\,$\mu$m band. 
In consideration of the uncertainties in the continuum determination, 
care was taken to use the same ``ice-free'' regions and 2-order polynomia for
all sources. Further discussion of the determination of continua
for extracting ice optical depth spectra can be found in e.g. \cite{Gerakines95, Dartois02, Keane01}. 
The ice spectra are shown on an optical depth scale in Fig. \ref{features}, while the 
locations of the sources relative to an 850\,$\mu$m JCMT-SCUBA map obtained by the COMPLETE collaboration \citep{Ridge06} are shown
in Fig. \ref{core_map}. 

 The CO ice column densities have been determined in \cite{Pontoppidan03a}. For the water 3.08\,$\mu$m band and the CO$_2$ 
15.2\,$\mu$m band, band strengths of $2.0\times 10^{-16}\,\rm cm\,molecule^{-1}$ and $1.1\times10^{-17}\,\rm cm\,molecule^{-1}$ are used, 
respectively \citep{Gerakines95}. For the 6.8\,$\mu$m band, a band strength of $4.4\times 10^{-17}\,\rm cm/\,molecule^{-1}$ is assumed, 
appropriate for NH$_4^+$ \citep{Schutte03}.
The ice column densities are summarized in Table \ref{parameters}.

\begin{table*}
\footnotesize
\centering
\begin{flushleft}
\caption{Line-of-sight averaged ice abundances in the Oph-F core}
\begin{tabular}{llllll}
\hline
\hline
&CBRR 2422.8-3423 & IRS 43& IRS 42& IRS 44 & IRS 46\\
\hline
Distance to Oph-F MM2 [AU]&1.8e3&4.3e3&8.2e3&11.0e3&13.9e3\\
N(H$_2$O) [$\rm cm^{-2}$] &3.9e18&2.96e18&1.87e18&3.04e18&1.17e18\\
N(Pure CO)/N(H$_2$O)&0.78&0.32&0.15&0.06&0.05\\
N(CO in water)/N(H$_2$O)&0.19&0.073&0.05&0.04&0.0\\
N(CO$_2$)/N(H$_2$O) &0.32&0.24&0.184&0.154&0.103\\
N(6.8)/N(H$_2$O)$^1$ &0.08&0.10&--&0.12&--\\
 &ISAAC/IRS&ISAAC/ISOCAM&ISAAC/ISOCAM&ISAAC/ISOCAM&ISAAC/IRS\\
\hline
\end{tabular}
\label{parameters}
\end{flushleft}
\end{table*}

\section{Abundance profiles of CO and CO$_2$ ice}
\label{Icemap}
\begin{figure}
\centering
\includegraphics[width=7.cm]{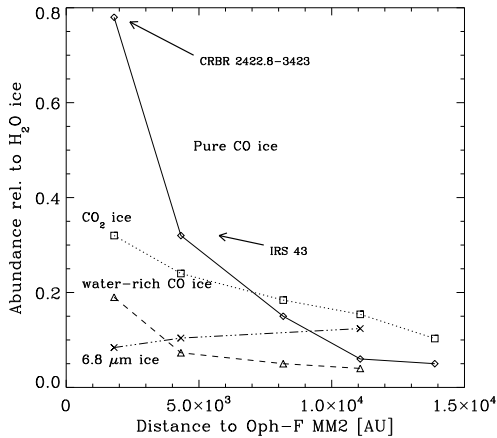}
  \caption{Radial map of CO and CO$_2$ ices of the Oph-F core. Ice 
abundances relative to H$_2$O ice toward young stars are plotted as functions of projected distance to the center of the core as 
described in the text. The CO ice has been split into a ``pure'' and ``water-rich'' component following \cite{Pontoppidan03a}.
}
  \label{COicemap}
\end{figure}

The basic ice map is constructed by determining the ice column densities along each line of sight and ratioing with the water ice column
density. The implicit assumption is that the 3.08\,$\mu$m water ice band traces the total H$_2$ column density (see \cite{Pontoppidan04} for a discussion).  
This yields CO and CO$_2$ line-of-sight averaged ice abundances relative to H$_2$O ice as a function
of projected distance to the center of the core. The radial map is shown in Fig. \ref{COicemap}. 

It is seen that the abundances of pure CO, 
water-rich CO and CO$_2$ all increase toward the center of the core, while the 6.8\,$\mu$m band carrier decreases slightly in abundance. 
The sharp rise in CO ice abundance is expected for CO freezing 
out from the gas-phase at high densities in the central parts of the core. 
Assuming a constant water ice abundance of $9\times 10^{-5}$ relative to H$_2$ in accordance 
with a model of the CRBR 2422.8-3423 line of sight of \cite{Pontoppidan05}, the average abundance of pure CO ice is seen to rise from 
$4.5\times 10^{-6}$ to $7.0\times 10^{-5}$, and the total CO and CO$_2$ abundances from $2.3\times 10^{-5}$ to 
$12\times 10^{-5}$. This corresponds to a total CO depletion ranging from 12\% to 60\%, assuming an initial CO gas-phase 
abundance of $2\times 10^{-4}$.

The increase by roughly a factor two in CO$_2$ ice abundance is particularly interesting. Since
CO$_2$ molecules are formed on the grain surfaces, the increase in CO$_2$ ice abundance along with the CO freeze out is a direct
indication of a formation route associated with CO. A significant amount of CO$_2$ is present in
the outer parts of the Oph-F core where the densities are not yet high enough for the CO to freeze out. 
{\it This can be interpreted as evidence for two distinct eras for CO$_2$ ice formation}: The first takes place at roughly the same time 
as the bulk of the water ice is formed, which is known to happen as soon as the extinction, $A_V$, into the cloud reaches a specific, 
relatively low, threshold of roughly 3-5 magnitudes \citep{Whittet88}. This domain is not probed by the Oph-F ice map. 
The second takes place during the catastrophic freeze-out of CO that occurs at
densities of a few $10^5\,\rm cm^{-3}$ \citep{Jorgensen05} in which an almost pure CO ice mantle forms. This second phase of CO$_2$
molecules should be easily detectable in the shape of the 15.2\,$\mu$m CO$_2$ bending mode absorption band, which is highly sensitive to the 
molecular environment of the CO$_2$ molecules. Specifically, the new CO$_2$ molecules should be found to be dilute in the CO ice 
with an average fraction
of 1:5, as determined by the relative CO and CO$_2$ abundance increases in the Oph-F core. Since the CO freeze-out rate is 
a very strongly increasing function of density, it is expected that the CO$_2$:CO ratio will decrease in denser parts of the core. Thus, 
one would expect to find a range of components in the CO$_2$ bending mode in the centers of dense clouds.
The spectral resolution of the ISOCAM-CVF spectra ($\lambda/\Delta\lambda\sim 50$) is not sufficiently high to search for such band shape
changes in the current data. 

Since the young stars that are used to estimate ice abundances in the core
material are embedded in the core, significant caveats apply. Heating of the core material may desorb CO ice within a radius of a 
few hundred AU of each source, depending on the luminosity. However, 
since the projected size of the core is $30\,000\,$AU, desorption is likely to play only a minor role. A possible exception
to this is IRS 46, which contains a significant component of warm molecular gas along the line of sight \citep{Lahuis06}. 
Also, the sources may be surrounded by remnant envelopes with different ice abundances than the surrounding core.
The ice absorption toward CRBR 2422.8-3423 was shown by \cite{Pontoppidan05} to be dominated by cold core material, although a fraction
of the water, CO$_2$ and 6.8\,$\mu$m ices are likely located in the disk. However, it
was found that the CO ice observed in this line of sight could not originate in the disk.  

\section{Empirical determination of the CO-CO binding energy}

The CO depletion profile in Fig. \ref{COicemap} is modeled using a simplified freeze-out model, assuming a static core and 
that only thermal desorption occurs. For the purposes of this letter, the model presented is intended as a demonstration of
the method, rather than achieving very accurate results. The rate of ice mantle build-up is then given by:

\begin{equation}
\frac{{\rm d}n_{\rm ice}}{{\rm d}t} = R_{\rm ads}-R_{\rm des},
\end{equation} 

Where $R_{\rm des}=\nu_0 \exp[-dH/kT]\times n_{\rm CO,ice}\times \beta$ is the desorption rate and 
$R_{\rm ads} = n_{\rm CO,gas}\times n_{\rm dust}\times \pi d^2\times \sqrt{3kT/m_{\rm CO}}\times f$ is the adsorption rate.
$\beta$ is a factor taking into account that CO only desorbs from the top monolayer (i.e. 1st order desorption for 
sub-monolayer coverage and 0th order desorption for multilayers). $d=0.05\,\mu$m is the grain radius, $f=1$ is the sticking coefficient, $\nu_0$
is the frequency of the CO stretching mode and
$n_{\rm CO,gas}$ and $n_{\rm dust}$ are the number densities of CO gas and dust particles. Solving for $n_{\rm ice}$ yields a time dependent
ice density given a gas density and (assumed identical) gas and dust temperature. This rate equation reaches an equilibrium on 
relatively short time scales ($<<10^6\,\rm years$). Therefore, given a temperature and density structure, one can in principle
solve for the binding energy of CO, $dH$. 
For the density profile of the core, a Bonnor-Ebert sphere with a central density of $5.5\times 10^{5}\,\rm cm^{-3}$ is used
with a temperature of 15\,K as suggested by \cite{Motte98} and references therein. This simple procedure yields a CO on CO binding energy 
of $814\pm 30$\,K. The uncertainty reflects different
results obtained if varying the central density and water ice abundance by 50\%. Additionally, the derived binding energy scales
linearly with the assumed dust temperature. Considering the simplified analysis, this result
is consistent with that recently found by \cite{Bisschop06} from laboratory experiments.

\begin{figure}
\centering
\includegraphics[width=7cm]{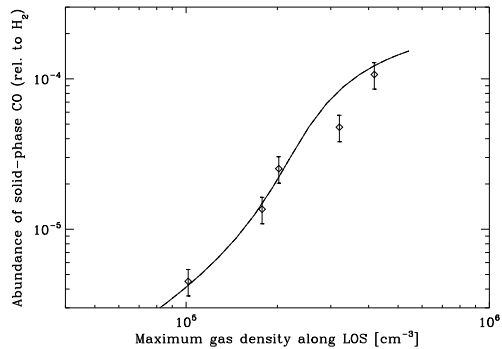}
  \caption{Observed and modeled CO ice abundance for Oph-F. The curve shows the simple
freeze-out model for the best-fitting CO binding energy of {$dH/k=814\,$K}. The ``solid CO'' abundance is the sum of the pure CO, CO in water
and the CO$_2$ ice excess abundances. This assumes that one CO molecule is used to produce each CO$_2$ molecule. The error bars indicate a 20\% uncertainty.}
  \label{COicemap_model}
\end{figure}

The radial map constitutes the first direct observation of the freeze-out profile of CO on dust grains in a prestellar core previously 
inferred indirectly from observations of millimetre lines of molecules. Additionally
the observed increase in CO$_2$ ice abundance toward the center is the first quantitative observational evidence of the formation
of CO$_2$ from CO on the surfaces of dust grains. These are observations that would not have been possible with single lines of sight.

\begin{acknowledgements}
      This work was supported by a Spinoza grant. The author gratefully acknowledges discussions with Helen Fraser and Ewine van Dishoeck. The
850\,$\mu$m map of Oph-F from the COMPLETE survey was supplied by Doug Johnstone.      
\end{acknowledgements}

\bibliographystyle{aa}
\bibliography{OphF}
\end{document}